\documentclass{aa}
\usepackage{graphics}

\begin{document}

   \title{VLBI astrometry of circumstellar OH masers;
        proper motions and parallaxes of four AGB stars}
   \titlerunning{VLBI astrometry of circumstellar OH masers}


   \author{W.H.T. Vlemmings\inst{1,}\inst{4}\and
        H.J. van Langevelde\inst{2}\and
        P.J. Diamond\inst{3}\and
        H.J. Habing\inst{4}\and
        R.T. Schilizzi\inst{2,}\inst{4}\
          }

   \offprints{WV (wouter@astro.cornell.edu)}

   \institute{Department of Astronomy, Cornell University, Ithaca, NY 14853-6801 
        \and
              Joint Institute for VLBI in Europe, Postbus 2, 
                7990~AA Dwingeloo
        \and
         Jodrell Bank Observatory, University of Manchester, Macclesfield,
                    Cheshire, SK11 9DL, England
	\and     
        Sterrewacht Leiden, Postbus 9513, 2300 RA Leiden, 
              the Netherlands}

   \date{Received ; accepted }


\abstract{ The main-line OH masers around 4 AGB stars have
  been observed with the NRAO Very Long Baseline Array (VLBA) at 8
  epochs over a period of 2.5 years. Using a phase referencing
  technique, the position of the most compact maser spot  of each star
  was monitored with respect to two extragalactic reference
  sources. For U~Her and W~Hya, we observe the most blue-shifted maser
  spot, while for R~Cas and S~CrB we only detect a compact red-shifted
  maser spot.  We managed to determine an accurate proper motion and
  parallax for U~Her, R~Cas and S~CrB, while additional motion of the
  compact blue-shifted maser of W~Hya is shown to possibly be related
  to the stellar pulsation. The motion and radio position are compared
  with the stellar trajectory and absolute optical position determined
  by the Hipparcos satellite. For U~Her and W~Hya, the most
  blue-shifted maser is consistent with the amplified stellar
  image. The new distances are compared with several published P--L
  relations, and in this respect the VLBI distances seem an
  improvement upon the Hipparcos distances.}

   \maketitle

\section{Introduction}

It was shown in van Langevelde et al. (2000, hereafter vL00) that
astrometric observations of the circumstellar OH masers can be used to
determine the proper motion and parallax of maser bearing stars. In
vL00, observations with the NRAO\footnote{The National Radio Astronomy
Observatory is a facility of the National Science Foundation operated
under cooperative agreement by Associated Universities, Inc.} Very
Long Baseline Array (VLBA) were performed on the 1665 and 1667~MHz
masers around the Mira variable star U~Her. The motion of the masers
was traced over 4 years and the proper motion and parallax of the
underlying star were determined.

In order to use the maser positions to monitor the stellar trajectory,
an assumption has to be made about the motion of the masers with
respect to the star. In vL00, it was confirmed that, in the case of
U~Her, the most blue-shifted circumstellar maser spot corresponds to
amplified emission originating from the stellar radiophotosphere.
The interpretation, in which a high brightness spot in the
blue-shifted peak of the OH maser spectrum is expected to mark the
line of sight to the star, will be referred to in this paper as the
{\it Amplified Stellar Image Paradigm}.  This interpretation was
already proposed by Norris et al. (1984). Very Long Baseline
Interferometry (VLBI) observations by Sivagnanam et al. (1990) also
provided strong evidence for this, as they showed that in U~Her the
dominant OH 1665 and 1667~MHz maser features at the blue-shifted side
of the maser shell were coinciding, in accordance with the
Paradigm. In vL00, the position of the compact blue-shifted maser
features was shown to fall directly onto the Hipparcos optical
position.  MERLIN observations in Vlemmings et al. (2002) indicated
that the amplified stellar image of U~Her is also seen at the 22~GHz
H$_2$O maser transition.

Maser astrometry yields direct distances to enshrouded
stars. This allows the inclusion of the more extreme Mira stars
in studies of the fundamental properties of these stars, like the
pulsation and mass-loss mechanism. The current discussion on the P--L
relation (e.g.\ Whitelock \& Feast 2000), for instance, could then
include stars with higher mass loss and generally longer periods.  At
the moment these investigations are based on Hipparcos distances,
which excludes the stars with high mass-loss because they tend to be
too obscured in the optical. A similar outstanding debate considers
the precise pulsation mode of Mira variables; this requires the
conversion of IR interferometry data into absolute diameters and hence
distances (van~Leeuwen et al.\ 1997; Wood\ 1998).

Here we observed an additional 3 stars in the astrometric OH
maser monitoring campaign which was also continued on U~Her. All
four stars were observed with Hipparcos (Perryman et al.  1997) and
the Hipparcos parallaxes were recently recalculated by Knapp et
al. (2003). However, because our target stars are faint and variable,
the Hipparcos results, especially on the parallax, are quite
uncertain. Our VLBI results improve upon the Hipparcos
results. They also further provide statistics on the {\it Amplified
Stellar Image  Paradigm}, confirming the results on U~Her and
W~Hya, while R~Cas and S~CrB surprisingly show only a compact
red-shifted maser spot that nonetheless could be traced for over two
years.

In \S 2 we discuss the observations and the absolute astrometry and
error analysis. In \S 3 we give the results of our parallax and proper
motion determination, which we discuss in \S 4. The conclusions are
presented in \S 5.

\section{Observations}

\begin{table*}
\caption{The sample}
\begin{tabular}{|l||c|c|c|c|rr|}
\hline
{ Source} & { Period} & $V_{\rm star} $ & { Hipparcos parallax} & Hipparcos Parallax $^a$&
\multicolumn{2}{c|}{{ Hipparcos Proper motion}} \\
& { (days)} & { (km~s$^{-1}$)} &
    { (mas)} & (mas) &
    \multicolumn{2}{c|} {{ RA, dec (mas/yr)}} \\
\hline
\hline
W~Hya & 361 & 40.0 & $8.73 \pm 1.09$ & $12.85 \pm 0.99$ & $-49.05 \pm 1.18$ & $-59.58$$\pm$$0.78$\\
S~CrB & 360 & 0.0 & $1.90 \pm 1.36$ & \hspace{3pt}$2.40 \pm 1.17$ & $-8.33 \pm 0.93$ & $-11.55$$\pm$$0.62$\\
U~Her & 406 & 14.5 & $1.64 \pm 1.31$ & \hspace{3pt}$1.88 \pm 1.31$ & $-16.84 \pm 0.82$ & $-9.83$$\pm$$0.92$\\
R~Cas & 430 & 26.0 & $9.37 \pm 1.10$ & $10.04 \pm 1.10$ & $84.39 \pm 0.95$ & $18.07$$\pm$$0.88$\\
\hline
\multicolumn{7}{l}{$^a$ recalculated by Knapp et al. (2003)}\\
\end{tabular}
\label{sam1}
\end{table*}

The positions of the 1665 and 1667~MHz circumstellar OH masers of a
sample of 4 AGB stars were monitored over a period of almost 2.5 years
with the NRAO VLBA. The sample of stars consisted of the Mira variable
stars R~Cas, S~CrB, U~Her and the Semi-Regular star W~Hya. The
observations were performed over 9 epochs. These are October 9
1999, January 20 2000, June 7 2000, September 1 2000, December 15
2000, March 11 2001, May 27 2001, February 23 2002 and May 12
2002. The observations of March 11 2000 failed due to heightened
ionospheric activity caused by the solar maximum. The OH
masers around U~Her were already previously monitored between July
1994 and April 1998 and the results of those observations are
presented in vL00.

The sources were selected to have bright 1665 and/or 1667~MHz OH
masers. Also, the sources were chosen so that two bright nearby phase
reference sources were available. Finally, the stars were selected to
be within $\approx 500$~pc in order to have detectable parallaxes of
more than $\sim 2$~mas. The stars in our sample with their period,
stellar velocity with respect to the Local Standard of Rest
(LSR) and Hipparcos proper motion and parallax are given in
Table~\ref{sam1}.

For each epoch the total observation time was 12 hours.  During the
first 2 epochs 4 additional stars were observed (R~Crt, RT~Vir, RS~Vir
and R~Aql). Both RS~Vir and R~Aql were observed at the 1612~MHz
instead of the 1665~MHz transition. R~Crt was not detected
and we were unable to get a phase connection to the phase reference
sources of RT~Vir, RS~Vir and R~Aql. We also found that in 12 hours we
can obtain only sufficient uv-coverage for 4 sources, because the
beam-shape is important to measure accurate positions.  For U~Her,
S~CrB and R~Cas this resulted in an average beam size of
$12\times7$~mas. Because of a low declination, the beam for W~Hya is
strongly elongated and is on average $20\times8$~mas.

\begin{table*}[t!]
\caption{Sources with corresponding reference sources}
\begin{tabular}{|l|c||l|l|l|c|r|}
\hline
Source & Ref. & \multicolumn{1}{c|}{Calibrator} & \multicolumn{1}{c|}{RA} & \multicolumn{1}{c|}{Dec} & Sep. & \multicolumn{1}{c|}{Flux} \\
& \# & & \multicolumn{1}{c|}{($^{h}~^{m}~^{s}$)} & \multicolumn{1}{c|}{($^\circ~{'}~{``}$)} & ($^\circ$) & \multicolumn{1}{c|}{(mJy)} \\
\hline
\hline
W~Hya & 1 & { J1339-262} & { 13~39~19.890747}
  & { -26~20~30.49590} & 2.9 & $515 \pm 62$\\
 & 2 & { J1342-290} & { 13~42~15.345608} &
      { -29~00~41.83114} & 1.6 & $162 \pm 44$ \\
\hline
S~CrB & 1 & { J1522+3144} & {
  15~22~09.991716} & \hspace{3pt}{ 31~44~14.38214} & 0.4 & $362 \pm 79$ \\
      & 2 & { J1527+3115} & { 15~27~18.73703}
      & \hspace{3pt}{ 31~15~24.38625} & 1.3 & $151 \pm 15$ \\
\hline
U~Her & 1 & { J1636+2112} & { 16~36~38.18373}
  & \hspace{3pt}{ 21~12~55.5991} & 3.5 & $217 \pm 30$\\
      & 2 & { J1630+2131} & { 16~30~11.23117}
      & \hspace{3pt}{ 21~31~34.3144} & 2.8 & $~92 \pm 12$ \\
\hline
R~Cas & 1 & { J2355+495} & { 23~55~09.458179}
  & \hspace{3pt}{ 49~50~08.34000} & 1.6 & $832 \pm 70$ \\
      & 2 & { J2347+5142} & { 23~47~04.83800}
      & \hspace{3pt}{ 51~42~17.87700} & 1.8 & $~72 \pm 12$ \\
\hline
\end{tabular}
\label{calt}
\end{table*}

In every star both the maser lines were observed in dual polarization, with a
bandwidth of $500$~kHz centered on the stellar velocity. They were
correlated with moderate spectral resolution ($1.95$~kHz$=0.36$
km~s$^{-1}$). Simultaneously two 4~MHz wide bands were recorded to
detect the continuum reference sources. These were correlated with a
spectral resolution of $31.7$~kHz for the final two epochs and
$125$~kHz on the others.

\begin{figure}[t!]
  \begin{center}
   \resizebox{\hsize}{!}{\includegraphics{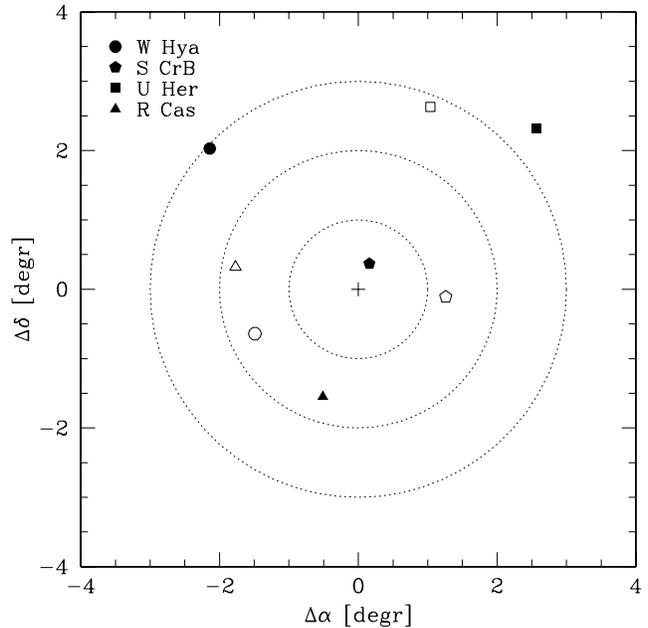}}
  \end{center}
\hfill \caption{
   Reference source positions wrt the target star in Right Ascension
   and Declination off-set. The sources are indicated with the listed
   symbols where the solid symbols are reference source 1, and the
   open symbols reference source 2. The circles are plotted to guide
   the eye and have a diameter of $2, 4$ and $6$ degrees.}
   \label{cal}
\end{figure}

The extragalactic reference sources were selected from the VLBI
calibrator catalogue (Beasley et al. 2002). The positions of the
reference sources for U~Her, J1636+2112 and J1630+2131 (formerly
J1628+214), were refined in vL00. In Table~\ref{calt} we list the
observed sources with their reference sources, and the separation
between the star and the reference source. Our phase reference cycle
was 5 minutes, with 3 minutes per source and then 1 minute for each of
the calibrators. The calibrator positions with respect to the source
position are plotted in Fig.~\ref{cal}.

The data was then processed in AIPS without any special astrometric
software. We rely on the VLBA correlator model and work with the
residual phases directly. To be able to apply the phase, delay and
phase rate solutions obtained on the continuum reference sources, a
special task was written to connect the calibration of the wide band
data to the spectral line data.

\subsection{Ionospheric effects}

\begin{figure}[ht!]
\begin{center}
   \resizebox{\hsize}{!}{\includegraphics{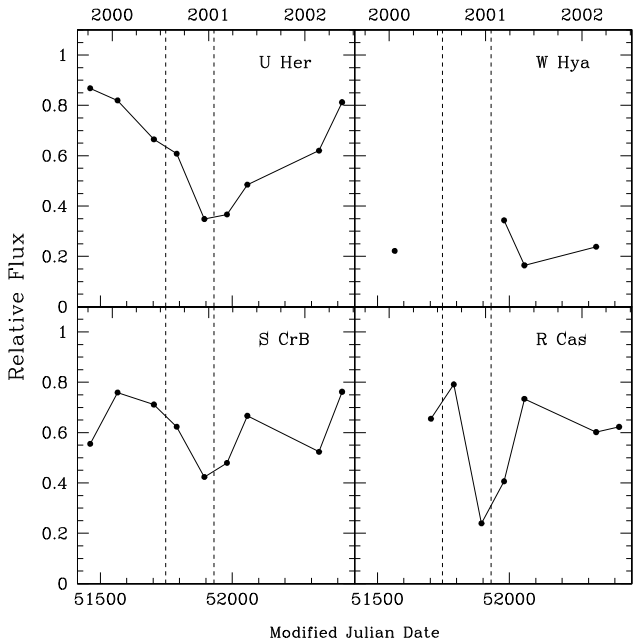}}
  \end{center}
 \hfill \caption{ The ratio between the flux density of reference source 1
 when phase-referenced to reference source 2 and the flux density when using
 the phase, delay and phase rate solutions obtained on reference
 source 1 itself. The dashed vertical lines indicates the approximate
 period maximum solar activity.}
   \label{relflux}
\end{figure}

 The correlator model does not include an ionospheric model. Phase
referencing will approximate the ionospheric conditions by zero order
in position and first order in time.  For our analysis no additional
ionospheric modeling is included. As noted in vL00, the remaining
effects of ionospheric activity can be seen as slightly distorted
images

 Unfortunately, the observations were performed during the period of
solar maximum, which occurred in the last months of 2000. Due to this
maximum in solar activity the ionosphere affects the phase
referencing, which is illustrated in Fig.~\ref{relflux}. Here we
display the ratio between the flux density of one of the reference sources
when it is imaged directly and the flux density when it is imaged with the
calibration solution on the second reference source. While for the
reference sources of W~Hya the sources were too far apart to obtain
good phase connection for most of the epochs, we notice that on the
other three pairs of reference sources a significant amount of
coherence on these scales is lost due to ionospheric activity.
Furthermore, we also find that the phase connections are slightly
worse during daytime. Thus, as a result of the heightened ionospheric
activity, the positional accuracy will be less.

\subsection{Absolute astrometry}

\begin{figure}[t]
\begin{center}
   \resizebox{\hsize}{!}{\includegraphics{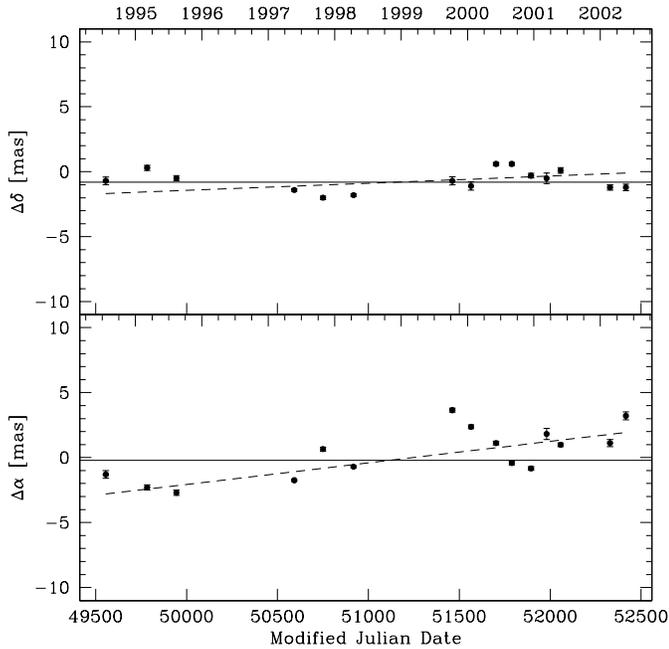}}
\end{center}
   \hfill \caption{The difference between the measured separation and
   its a-priori value of U~Her reference source 1 (J1636+2112) with
   respect to 2 (J1630+2131, formerly J1628+214). The solid line
   indicates the average separation between the reference sources. The
   dashed line is a weighted least square fit to a change in measured
   separation due to evolving source structure of J1630+2131.}
   \label{uherref}
\end{figure}

\begin{figure}[ht]
  \begin{center}
   \resizebox{\hsize}{!}{\includegraphics{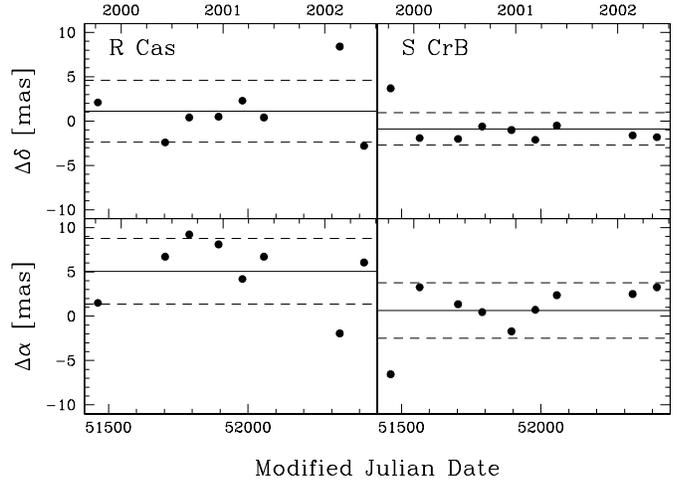}}
  \end{center}
   \hfill \caption{ The difference between the measured separation and
   its a-priori value of S~CrB reference source 1 (J1522+3144) with
   respect to 2 (J1527+3115), and R~Cas reference source 2
   (J2347+5142) with respect to 1 (J2355+495). The formal error bars lie
   within the plotted symbols. The solid line indicates the average
   separation between the reference sources and the dashed lines are
   the rms.}
   \label{2refs}
\end{figure}

The absolute astrometry of the maser spots was determined with respect
to the two extragalactic reference sources. As a consistency check we
also determined the positions of the reference sources with respect to
each other. The results for the reference sources of U~Her are shown
in Fig.~\ref{uherref}, and the results for those of S~CrB and R~Cas in
Fig.~\ref{2refs}. As was already seen in Fig.~\ref{relflux}, no
accurate phase connection was established between the reference
sources of W~Hya. We find that the reference sources J1630+1231 (of
U~Her) and J2355+495 (of R~Cas) show relatively large extended
structure on the VLBA baselines. J1630+1231 has some indication of a
single jet structure, while J2355+495 clearly shows two radio
lobes. For the first epoch of U~Her a detailed source model was used
for the phase referencing, but this seemed to have little effect.  For
the epochs presented here a point source model was used and the effect
on the astrometric positions is shown in Figs.~\ref{uherref} and
\ref{2refs}. When examining the separation between the reference
sources of U~Her, we notice that source evolution might cause a slight
drift in the obtained relative positions. A linear fit to this motion
indicates that the astrometric positions with respect to J1630+1231
show a drift of $\approx 0.57$ and $0.10$~mas/yr in right ascension
and declination respectively. This would influence the derived proper
motions by a similar amount, and it explains some of the discrepancy
between the Hipparcos and VLBI proper motions.  The determined parallax
will not be affected as no periodic motions correlated between right
ascension and declination seem to be present.  The
double lobe structure of J2355+495 does not seem to cause any
systematic effect, although the scatter on the position determination
is slightly bigger. However, this is likely also due to the fact that
the scatter in the difference between measured and a-priori assumed
separation increases with increasing source separation.

\subsection{Error analysis}

 There are several sources of errors in the determination of our
astrometric positions. One of these is the influence of the
ionosphere. As discussed above, the phase referencing will solve for
the zeroth order effect. The errors as a
result of the phase referencing over the ionosphere between the
reference source and the maser source are estimated by examining the
difference between the measured separation between the reference
sources and their a-priori expected separation as shown in
Figs.\ref{uherref} and \ref{2refs}.  From this we first determine the
average offset of the reference sources with respect to the expected
positions; this error is due to the error in the {\it absolute}
position of the reference sources.  Then we determine the scatter,
which can be attributed to the phase referencing. For R~Cas we find
that the difference between the measured separation and its a-priori
value is $5.1 \pm 3.6$~mas. For S~CrB we find $1.1 \pm 2.5$~mas. If
for U~Her, we simply take the average offset we find $0.71 \pm
1.5$~mas. When using the least-square fitted line discussed in the
previous section the scatter decreases to $1.1$~mas. Unfortunately, no
error analysis could be performed for W~Hya, and we take the
errors to be similar as for R~Cas.

Depending on the position of the target source with respect to the
reference sources, we generally find the largest scatter along the
declination axis. As the depth of the ionosphere along the line of
sight changes faster with a change in declination than with a change
in right ascension, the ionospheric errors in right ascension are less
than those in declination for a separation which is comparable in each
coordinate.

\begin{figure*}[ht!]
  \begin{center}
   \resizebox{\hsize}{!}{\rotatebox{-90}{\includegraphics{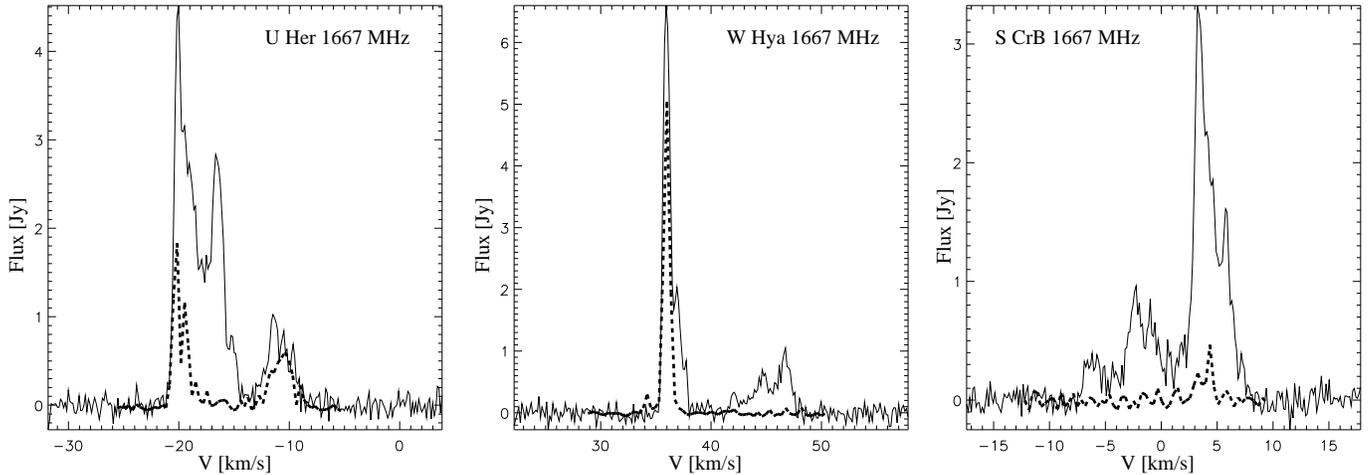}}}
  \end{center}
   \hfill \caption{The 1667~MHz maser spectra of U~Her, W~Hya and
   S~CrB. The solid line is the single dish spectrum in left-hand
   circular polarization for U~Her and S~CrB and right-hand circular
   polarization for W~Hya. The spectra are from Etoka \& Le Squeren
   (2000) and Etoka et al. (2001), and were taken in 1984 (U~Her) and
   1986 (W~Hya and S~CrB) at the Nancay radio telescope. The thick
   dashed line is the maser spectrum at a short VLBA baseline (Pt-La)
   from the last observing epoch (May 12 2002).}
   \label{3spec}
\end{figure*}

According to the VLBA calibrator list, the positions of all the
reference sources, except for J2347+5142 (from R~Cas) and J1630+2131
(from U~Her), are known within $1$~mas accuracy. In vL00 we have
observed J1630+2131 ourselves, and the positions used here are also
accurate within $1$~mas. The position of J2347+5142 is still only
known to within $20$~mas. Thus, the average positional offset for the
reference sources of U~Her and S~CrB agrees well with the estimated
accuracy of $\approx 2$~mas for the {\it absolute} positions of the
reference sources. The average offset for the calibrators of R~Cas is
larger, which was expected as the J2347+5142 positions are less
accurately known.

Errors also arise from the formal positional fitting. The
fitting was performed with the routine {\sc JMFIT} in AIPS, where we
fitted the 6 components of a Gaussian feature (position, long and
short axis, position angle and peak flux). The formal errors depend on
the beam size and the thermal fluctuation in the mapped spectral
channel where the maser feature occurs. The errors are approximately
proportional to ${\rm Beamsize} / {\rm SNR}$. This is generally less
than $1$~mas for most of our observations, although, since the
signal-to-noise ratio (SNR) is also affected by the ionosphere, the
error could reach $\approx 3$~mas in each coordinate for W~Hya and
$\approx 1 - 1.5$~mas for the other sources. As most of our maser
features are resolved, higher resolution will not increase the
positional accuracy.

\section{Results}

\begin{figure*} 
   \resizebox{12cm}{!}{\includegraphics{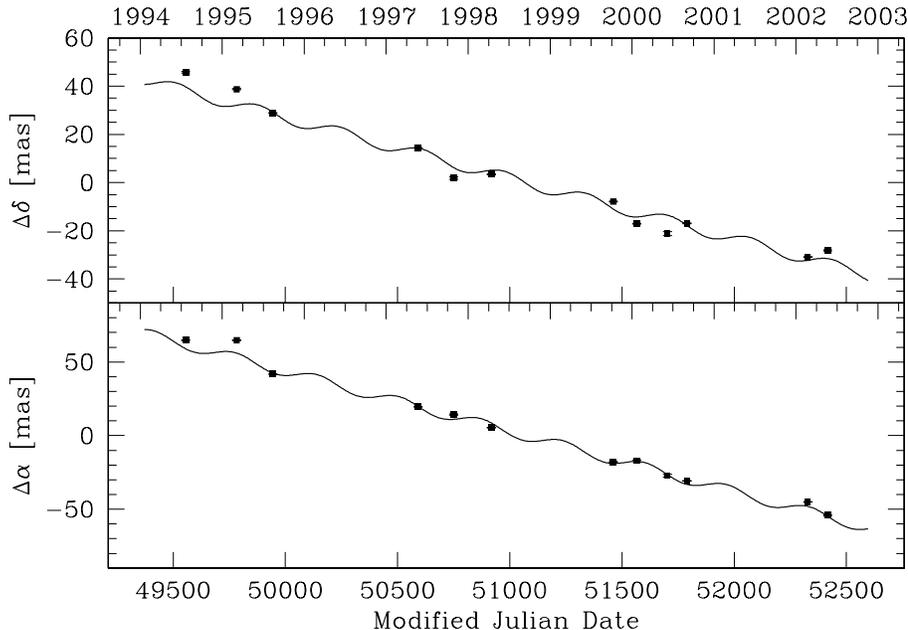}} \hfill
   \parbox[b]{55mm} { \caption{ The position of the most blue-shifted
   1667 MHz maser spot of U~Her with respect to J1630+2131. The error
   bars on the positions indicate the formal position fitting
   errors. The solid line is the best fitting parallax and proper
   motion trajectory for the 12 epochs of observations. Note that the
   scales are different for declination and right ascension.}
   \label{uherres} }
\end{figure*}

\subsection{Morphology}

 Of our sources, only U~Her shows multiple blue and red-shifted
features on VLBA baselines. We only managed to identify one
bright maser feature for S~CrB, W~Hya and R~Cas. We were able to image
both 1665 and 1667~MHz masers in U~Her, W~Hya and S~CrB, while for
R~Cas only the 1665~MHz maser was bright enough.

 Fig.~\ref{3spec} shows the 1667 spectrum of U~Her on one of the short
VLBA baselines compared with the single dish spectrum. We see that the
features span a range between $-8$ and $-21$~km~s$^{-1}$ around the stellar
velocity of $-14.5$~km~s$^{-1}$. Due to the ionospheric activity the double
peaked spectrum was only observed in the epochs presented in vL00, and
in the last two epochs. For the other epochs we could only map
the most blue-shifted feature. Thus, we did not include other
maser spots in our astrometric analysis. At the 1665~MHz transition,
the most blue shifted spot is found slightly shifted in velocity with
respect to the 1667~MHz most blue-shifted feature. The most
blue-shifted 1665 feature has a velocity of $-20.8$~km~s$^{-1}$, while the
most blue-shifted 1667 feature has a velocity of
$-20.4$~km~s$^{-1}$. Similarly as in vL00, we find that in most cases the
bright 1665 and 1667 maser features coincide within the synthesized
beam. Since at the epochs in vL00 we only used the most blue-shifted
1667 maser spot to determine the proper motions and parallax, here we
also only use the 1667 spot.

 In W~Hya we only detected one bright maser spot at $35.6$~km~s$^{-1}$ in
both maser transitions.  Fig.~\ref{3spec} shows a short VLBA baseline
1667 spectrum compared with the single dish spectrum. The positions of
the 1665 and 1667 maser spots coincide within the beam, which,
since W~Hya is observed at very low declination, is highly elongated
along the declination axis. The bright feature we detected corresponds
in velocity to the brightest maser feature detected by Szymczak et
al. (1998). This is the most blue-shifted feature at 1665~MHz, whereas
at 1667~MHz a more blue-shifted, slightly weaker feature is found at
$33.9$~km~s$^{-1}$. This weaker feature was only detected at our last epoch
and is found to exist at $\approx 17.2$~mas from the brightest feature.

 Around S~CrB we detected only one bright maser feature at both
transitions, which is red-shifted with respect to the stellar velocity
of $0.0$~km~s$^{-1}$. The feature at 1665~MHz is found to coincide within the
beam with the 1667~MHz spot, although the velocity at 1667~MHz is
$3.2$~km~s$^{-1}$ and the velocity at 1665~MHz is $2.9$~km~s$^{-1}$. We find that
these features correspond to the brightest features in the single dish
spectra presented by Etoka \& Le Squeren (2000) and the spectrum seen
in Fig.~\ref{3spec}, which is not necessarily the most
red-shifted. Also in the single dish spectra, the main-line masers
show only weak blue-shifted emission, while the 1612~MHz satellite
line shows a strong blue-shifted peak.

 Finally, around R~Cas we only detected a narrow maser feature at
1665~MHz, at a velocity of $29.5$~km~s$^{-1}$. This is red-shifted with
respect to the stellar velocity of $26.5$~km~s$^{-1}$. In the single dish
spectrum observed by Chapman et al.(1994), the brightest feature is
found red-shifted as well, at a velocity of $29.2$~km~s$^{-1}$. However, this
was not the most red-shifted feature, as still weaker features were
found at higher velocity.

\begin{table*}
\caption{Results}
\begin{tabular}{|l|c||c|rr|c|}
\hline
Source & MHz & VLBI parallax & \multicolumn{2}{c|}{VLBI Proper motion} & \# epochs \\
&  & (mas) & \multicolumn{2}{c|} {RA, dec (mas/yr)} & \\
\hline
\hline
W~Hya & 1665 & $9.79 \pm 3.23$ & $-45.54 \pm 3.02$ & $-53.82  \pm  5.08$ & 8\\
      & 1667 & $12.50 \pm 4.23$ & $-41.80 \pm 3.14$ & $-56.39  \pm  3.65$ & 8\\
      & both & $10.18 \pm 2.36$ & $-44.24 \pm 2.04$ & $-55.28  \pm  2.93$ & \\
\hline
S~CrB & 1665 & $2.37 \pm 0.43$ & $-8.92 \pm 0.38$ & $-12.21 \pm 0.65$ & 7 \\
      & 1667 & $2.30 \pm 0.50$ & $-9.13 \pm 0.41$ & $-12.53 \pm 0.46$ & 8 \\
      & both & $2.31 \pm 0.33$ & $-9.08 \pm 0.27$ & $-12.49 \pm 0.33$ & \\
\hline
U~Her & 1667 & $3.61 \pm 1.04$ & $-14.94 \pm 0.38$ & $-9.17 \pm 0.42$ & 12 \\
\hline
R~Cas & 1665 & $5.67 \pm 1.95$ & $80.52 \pm 2.35$ & $17.10 \pm 1.75$ & 8\\
\hline
\end{tabular}
\label{res1}
\end{table*}

\subsection{Proper motion / parallax}

 The proper motion and parallax for our sample of sources were
determined using a least-square fitting method. For U~Her we included
the data of the 6 epochs presented earlier in vL00. When the spot was
observed in both transitions (for W~Hya and S~CrB), we performed the
fit on both the spots separately as well as on the combined data.
The errors due to the thermal fluctuation are not correlated for the
two frequencies, but the systematic ionospheric effects can be. The
results of the fits are presented in Table\ref{res1} along with the
number of epochs at which the maser feature was detected. The errors
represent the $1\sigma$ deviations.  In the further analysis for
W~Hya and S~CrB, we take the fit to the combined 1665 and 1667 data as
the most reliable estimate.

\begin{figure}
   \resizebox{\hsize}{!}{\includegraphics{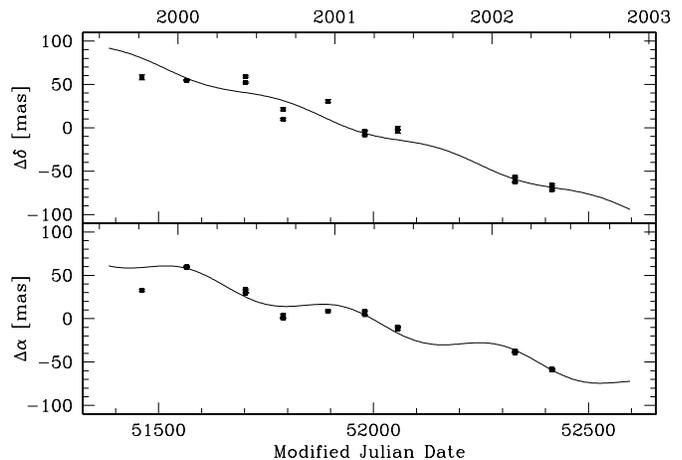}}
   \hfill \caption{ Similar to Fig.~\ref{uherres}. The position of the
   most blue-shifted 1665 and 1667 MHz maser spots of W~Hya with
   respect to J1342-290. The 1665 and 1667 maser spots have the same
   velocity. Drawn is the best fitting parallax and proper motion
   trajectory combining the results on both maser transitions for the
   8 epochs of observations.}
   \label{whyares}
\end{figure}

\begin{figure}
   \resizebox{\hsize}{!}{\includegraphics{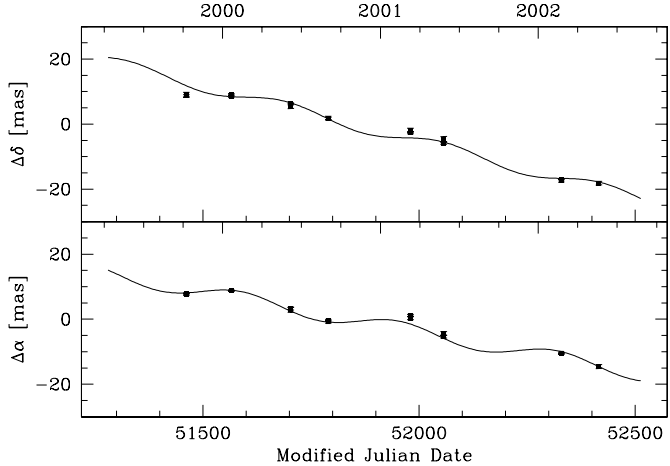}}
   \hfill \caption{ Similar to Fig.~\ref{uherres}. The position of the
   brightest 1665 and 1667 MHz maser spots of S~CrB with respect to
   J1522+3144. The brightest S~CrB maser feature is red-shifted with
   respect to the stellar velocity. The 1665 and 1667 maser spots have
   the same velocity. Drawn is the best fitting parallax and proper
   motion trajectory combining the results on both maser transitions
   for the 8 epochs of observations.}
   \label{scrbres}
\end{figure}

\begin{figure}
   \resizebox{\hsize}{!}{\includegraphics{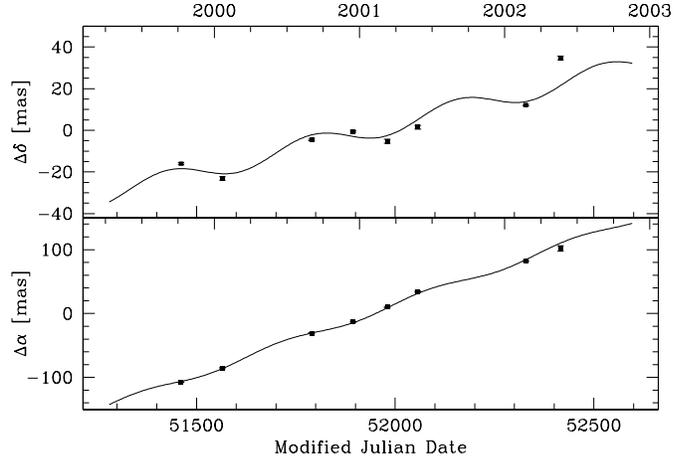}}
   \hfill \caption{ Similar to Fig.~\ref{uherres}. The position of
   the brightest 1665 MHz maser spot of R~Cas with respect to
   J2355+495. The brightest R~Cas maser feature is red-shifted with
   respect to the stellar velocity. Drawn is the best fitting parallax
   and proper motion trajectory combining the results on both maser
   transitions for the 8 epochs of observations.}
   \label{rcasref}
\end{figure}

 The fitted motions and observed positions are presented in
Figs.~\ref{uherres}, \ref{whyares}, \ref{scrbres} and
\ref{rcasref}. The errors displayed are the formal uncertainties in
the Gaussian profile fitting used to determine the positions. As
discussed above these were less than $1$~mas at most epochs, but could
go up to $\approx 1.5$~mas (or even $3.0$~mas in the case of W~Hya) as
the ionospheric conditions worsened.

As previously seen in vL00, the final fit for U~Her leaves
rms residuals somewhat larger than those estimated to be caused by the
phase referencing. These cannot be caused by the possible linear drift
of the reference source position, as this will only change the
measured proper motion. We find weighted rms residuals of $3.0$ and
$3.1$~mas in right ascension and declination respectively, while from
the reference sources we expected residuals of $\approx 1.5$~mas. It
should be noted however, that the separation between U~Her and its
reference source is bigger than the separation between the reference
sources themselves, and this we think is the cause of the larger
residuals.  The separation between U~Her and its reference source is
$2.8^\circ$, similar to the separation between the R~Cas reference
sources. Thus residuals of $\approx 3$~mas can be expected.

For S~CrB we find weighted rms residuals of only $1.0$ and $1.3$~mas
in right ascension and declination. This agrees very well with the
systematic errors in the relative astrometry originating from the
ionospheric activity, which were estimated to be $\approx
2.5$~mas. The residuals are smaller likely due to the fact that the
source-reference source separation is less than between the reference
sources. The weighted rms residuals for R~Cas are somewhat larger
($2.2$ and $4.6$~mas), but, when the position fitting errors are taken
into account, they still agree with the expected value of
$3.6$~mas. Finally, the weighted rms residuals for W~Hya are quite
large ($7.8$ and $10.2$~mas). Although we were unable to examine the
ionospheric effects on the reference sources, these residuals are too
large to be caused only by ionospheric effects. The maser spots seem
to show some additional systematic periodic motion, which is discussed in \S 4.1.2.

 For both U~Her and W~Hya, where the most blue-shifted maser spot is
expected to be the amplified stellar image, we have also performed a
fit including the Hipparcos optical positions as an additional data
point. From the Hipparcos catalogue we take the positional uncertainty
to be $0.1$~mas. The fit on U~Her then resulted in $\mu = -14.86 \pm
0.25, -9.50 \pm 0.27$~mas/yr and $\pi = 3.34 \pm 0.90$. The fit on
W~Hya gave $\mu = -51.69 \pm 0.28, -62.00 \pm 0.46$~mas/yr and $\pi =
9.65 \pm 3.00$~mas. However, one should realize that using the
Hipparcos position as an additional data point encompasses one more
uncertainty, namely the connection between the optical and radio
reference frames, which in principal introduces additional systematic
errors.

\subsection{Hipparcos Comparison}

\begin{table}[t!]
\caption{Hipparcos Flags}
\begin{center}
\begin{tabular}{|l||c|c|c|c|}
\hline
Source & H29 & H30 & H59 & H61 \\
 & (\%) &  &  &   \\
\hline
\hline
W~Hya & 4 & 5.15 & V &  \\
S~CrB & 2 & 2.31 & V &  \\
U~Her & 0 & 2.24 &   & S  \\
R~Cas & 0 & 2.53 & V &  \\
\hline
\end{tabular}
  \end{center}
\label{hipflag}
\end{table}

By comparing Table~\ref{res1} with Table~\ref{sam1}, we find that the
proper motions determined from the VLBI observations agree marginally
within the errors with the proper motions obtained with the Hipparcos
satellite. The Hipparcos parallaxes for U~Her and S~CrB were quite
uncertain, although they still agree with the VLBI parallaxes. Also
the W~Hya Hipparcos parallax agrees with our observations. The R~Cas
VLBI parallax is however, significantly smaller than found with
Hipparcos. The Hipparcos parallaxes that were recalculated by
Knapp et al. (2003) slightly improve the comparison for W~Hya, S~CrB
and U~Her, but the errors for S~CrB and U~Her remain large. The R~Cas
VLBI parallax remains significantly smaller.

It is important to realize that, because the stars in our sample are
relatively faint variable AGB stars, some flags were attached to the
Hipparcos catalogue entries regarding the processing of the astrometric
data.  The flags relevant to our comparison are listed in
Table~\ref{hipflag}.  The H29 flag gives the percentage of data that
had to be rejected in order to obtain an acceptable astrometric
solution. Both W~Hya and S~Crb had some data rejected, most likely due
to observations at minimum light. The H30 flag indicates the
goodness-of-fit (F2) of the astrometric solution. It is stated in the
catalogue that fits with an F2 value exceeding +3 indicate a bad fit to
the data. We immediately see that the Hipparcos result for W~Hya is a
very bad fit, while also the other 3 stars give only moderately
acceptable fits. H59 is the flag indicating a possible double or
multiple system, where 'V' are labeled as 'variability induced
movers'. This indicates that as a result of variability the data show
additional motion. In itself this does not indicate bad astrometry, as
it is simply a label given to many strongly variable stars. However, it
does indicate that the reliability of the astrometric fit could be
less. Finally an H61 flag 'S' indicates a suspected non-single system
when no significant or convincing non-single star solution is found.
However, the catalogue suggests that many sources labeled 'S' are
actually single stars, with the flag induced by variability or inadequate
sampling. The flags are more thoroughly discussed in Perryman et al.
(1997).

\begin{figure}[t!]
\begin{center}
   \resizebox{\hsize}{!}{\includegraphics{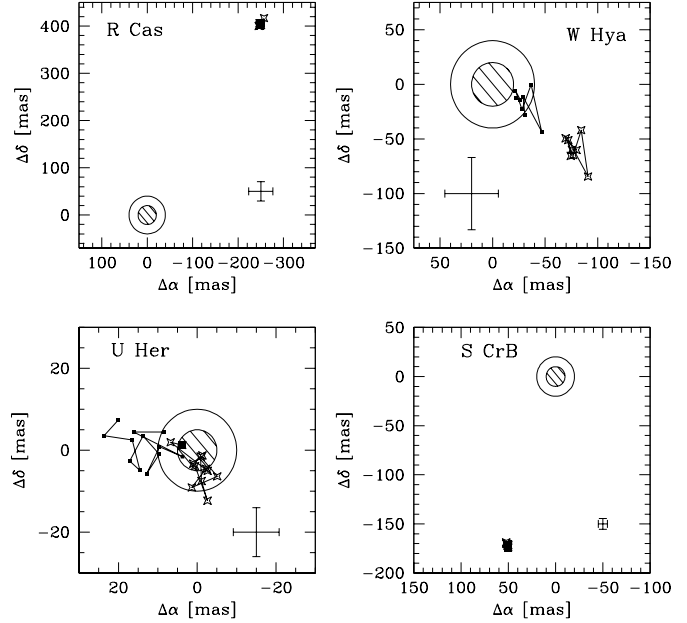}}
\end{center}
   \hfill \caption{The position of the observed maser spot wrt the
   optical position as predicted by our fits. The solid symbols denote
   the first epoch and subsequent epochs are connected. Also drawn are
   indications of the size of the star (shaded) and the
   radiophotosphere. The error bars in the corner are the errors due
   to the transposition of the radio positions and the Hipparcos
   position errors. The exact size of the error bars depends on the
   epoch of observations, the drawn error bars are for the middle
   epoch. For U~Her and W~Hya the squares indicate the transposed
   positions using the Hipparcos proper motion and parallax. Note the
   difference in scale of the four plots.}
   \label{hippos}
\end{figure}

Because of these flags, the Hipparcos values, and especially the quoted
errors should be treated with care. Especially for W~Hya these can
explain the differences between the Hipparcos and VLBI results.

The absolute astrometry of the observed maser spot positions can be
compared with the Hipparcos optical position transposed from its mean
observation epoch J1991.25 using the fitted proper motion. The errors
in the comparison are due to the errors in the proper motion (and
parallax) used to transpose the optical position to a common
epoch. The alignment of the Hipparcos frame with the International
Celestial Reference Frame (ICRF) was made at $\la 1$~mas accuracy
(Lestrade et al. 1995).  The results are shown in Fig.~\ref{hippos}.
The U~Her radiophotosphere was estimated have a diameter of $\approx
20$~mas in vL00. W~Hya was observed by Reid \& Menten (1997) to have a
radiophotosphere diameter of $\approx 80$~mas at 22~GHz. From
optical interferometry the diameter of R~Cas was estimated to be
$\approx 40$~mas, giving a radiophotosphere of $\approx 80$~mas if the
radiophotosphere is twice the size of the star. For S~CrB we have
simply drawn an average size of $40$~mas.

We find that the off-set between the maser and stellar positions is
largest for the observed red-shifted maser spots of R~Cas and
S~CrB. While the error of the transposed positions is $\approx 34$~mas
due to the extrapolation to a common epoch, we find a weighted average
off-set of $474.2$~mas with an rms scatter of $3.6$~mas for the maser
of R~Cas. At the assumed distance of R~Cas this corresponds to
$\approx 83$~AU. For S~CrB we find a weighted average off-set of
$178.5$~mas with an rms scatter of $1.2$~mas, while the error due to
the transposed positions is $\approx 7$~mas. This off-set corresponds
to $\approx 77$~AU at $433$~pc.

For the most blue-shifted maser spots of U~Her we found in vL00, that
the position matches the transposed Hipparcos optical position within
the errors. With the improved parallax and proper motion fit we find a
weighted average off-set of $4.1 \pm 3.1$~mas, corresponding to
$\approx 1.1 \pm 0.8$~AU at the distance of U~Her. With the errors of
the transposition being $\approx 8$~mas at the middle epoch, the most
blue-shifted maser clearly falls onto the stellar radiophotosphere
with a $\approx 10$~mas radius. For a comparison, the Hipparcos
optical position was also transposed using the Hipparcos values for
proper motion and parallax. Using these the off-set increases to $13.0
\pm 5.0$~mas.

Although the fit of the motion of W~Hya has large errors and left some
unexplained residuals, a similar a\-na\-ly\-sis was performed. Using our
fitted motion, we find a weighted average off-set of $99.4 \pm
9.4$~mas, which would correspond to $\approx 9.7$~AU. The errors due
to the transposition were $\approx 42$~mas due to the large error of
the proper motion. We thus find that the most blue-shifted maser spot
falls just outside the stellar radio-sphere with an assumed diameter
of $\approx 80$~mas. However, using the Hipparcos proper motion, the
off-set decreases to $36.2 \pm 11.6$~mas, clearly on the stellar
photosphere.

\section{Discussion}

\subsection{Stellar motion and maser morphology}

\subsubsection{U~Her}

The masers of U~Her were observed with respect to the reference source
J1630+2131, which is located at $2.8^\circ$ from U~Her. The absolute
astrometry was estimated to be accurate to $\approx 1.5$~mas as
determined from the relative reference source positions, but, as the
reference sources have a smaller separation than their distance to
U~Her, the errors on U~Her could be as large as 3 mas.

We managed to trace the most blue-shifted 1667~MHz OH maser spot of
U~Her for an additional 6 epochs, which combined with the observations
in vL00 gives an improved fit of the proper motion and parallax of
U~Her. The proper motion in declination and the parallax are
consistent with the results in vL00, but the proper motion in right
ascension is somewhat smaller. Using the new values, the maser
positions fall directly onto the transposed Hipparcos positions,
without the seemingly systematic off-set detected in vL00. The
separation between expected optical position and the radio position is
only $\approx 1.1$~AU and well within the diameter of the
radiophotosphere. The rms residuals are consistent with the rather large
target -- reference source separation.  This, together with possible
turbulence of $\approx 0.7$~mas/yr, can explain the residuals found in
vL00 and observed here. As
discussed above, the reference sources used to get the positions of
the U~Her maser spots shows some indication of source evolution, which
was determined to possibly increase the measured proper motion by
approximately $0.57$ and $0.10$~mas/yr in right ascension and
declination respectively. This does not affect the observed alignment
of the maser positions with the optical position in Fig.~\ref{hippos},
because the Hipparcos position, transposed using the fitted values,
would experience the same positional shift as the maser positions.

In the course of the observations we did not find any other maser
spots that were bright or persistent enough to trace its motion. The
most blue-shifted spot was consistently the brightest, unlike in one
of the epochs observed in vL00,

\subsubsection{W~Hya}
\label{whyasec}

\begin{figure}[t!]
\begin{center}
   \resizebox{\hsize}{!}{\includegraphics{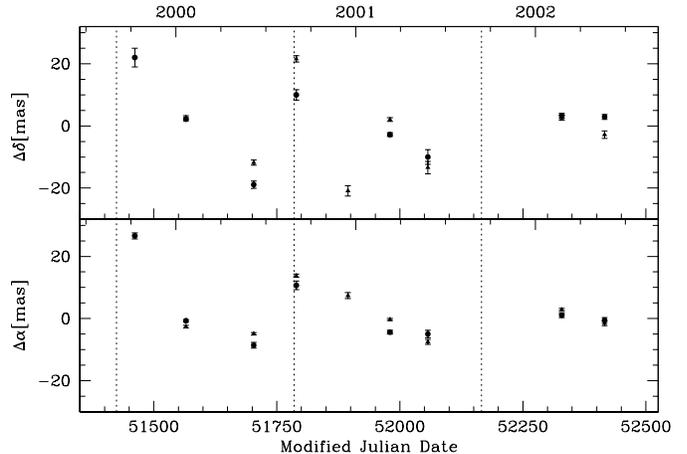}}
 \end{center}
  \hfill \caption{The residual motion of the maser spots of W~Hya
   after subtracting the best fitted model. The circles are the
   1667~MHz spot and the triangles the 1665~MHz spot. The errors are the
   formal position fitting errors. The dashed vertical lines indicate
   the position of the maximum of the stellar phase.}
   \label{whya_res}
\end{figure}

 The masers around the low declination source W~Hya were observed with
respect to the reference source J1342-290,  located at $1.6^\circ$
from W~Hya. We were unable to establish a phase connection between
the two reference sources themselves, but estimate the errors due to
the ionosphere to be $\approx 4$~mas.

 Also for W~Hya, we managed to trace the motion of the most
blue-shifted maser feature. It was detected at both transitions, and
the positions of the pair, matched to within $\approx 15$~mas,
approximately the size of the highly elongated beam. The best fitted
model has large errors and it has been shown that also the Hipparcos
results were labeled as a bad fit.  Using the best fitted models we
find that the maser spot position is off-set from the Hipparcos
optical positions by almost $10$~AU, however, when using the Hipparcos
proper motion the maser spots are only off-set by $\approx
3.5$~AU. Because the errors are large ($\approx 4$~AU), and the
assumed radius of the radiophotosphere is $\approx 4$~AU, we conclude
that also for W~Hya, the most blue-shifted maser spot is the amplified
stellar image. The residuals from our fit, shown in Fig~\ref{whya_res}
seem to describe some periodic fluctuations which are out of phase
with the motion due to the parallax. The residual motions have a
period between $300$ and $400$ days, and an amplitude of $\approx
15$~mas which at the determined distance of W~Hya corresponds to only
$\approx 1.5$~AU.  At the assumed distance of W~Hya ($\approx
100$~pc), a typical value of $1$~km~s$^{-1}$ turbulent motion in the maser
medium (Diamond et al. 1985) corresponds to $2.1$~mas/yr.  Therefore,
turbulence alone cannot explain the large residuals and it also seems
unlikely that the phase referencing could introduce such large
scatter. One possibility is that the stellar trajectory shows the
effect of a binary or multiple system. However, this can be ruled out,
as a motion of $15$~mas and a period of approximately a year would
indicate a companion of several solar masses within the stellar
diameter. Another cause of the residuals could be stellar
pulsations. There seems to be an indication that the residuals are
correlated with the optical light curve. Because the compact maser
spot is a result of strong beaming, the motions could result
from minor changes in the maser medium due to a variation in
pumping. Additionally, the maser spot can drift over the
radiophotosphere due to intensity variations in the stellar radio
emission, possibly caused by stellar pulsation. From optical
measurements it has been shown that between optical minimum and
maximum, the stellar diameter decreases by $\approx 20$~mas (Lattanzi
et al. 1997; Haniff et al. 1995). Lattanzi et al. have also shown that
W~Hya is highly elongated, with a major axis $20 \%$ larger than the
minor axis. Consequently, the stellar pulsation can easily explain the
large residuals observed.

\subsubsection{R~Cas}

 The masers around R~Cas were observed with respect to the reference
source J2355+495, at $1.6^\circ$ from the star. The errors due to the
ionosphere were estimated to be $\approx 3.6$~mas.

 The motion of R~Cas was determined from a fit to the positions of a
compact red-shifted maser spot at 1665~MHz. Since this spot is
therefore not fixed to the stellar radiophotosphere we have to assume
that the spot is stationary with respect to the star or on a linear
path. A turbulent motion of $1$~km~s$^{-1}$ at the distance of R~Cas would
result in additional motions of $1.5 - 2$~mas/yr. Additionally, we can
estimate the position shift we would observe due to a simple spherical
outflow. We assume the transposed Hipparcos position to be the stellar
position and the maser spot to be at $83$~AU (Fig.~\ref{hippos}), we
take the expansion velocity to be $5.5$~km~s$^{-1}$ and the OH maser extent
to be $\approx 300$~mas (Chapman et al. 1994). The maser spot would
then move outward from the star at $\approx 2.5$~mas/yr. This motion
does not affect the parallax, but can explain the difference between
the VLBI and Hipparcos proper motion. A parallax as large as the one
found by Hipparcos does not fit our data and taking all errors into
account we favor the VLBI parallax.

We can compare the location of the maser spot with respect to the star
in Fig.~\ref{hippos}, with the map of the 1665 and 1667 MHz OH masers
features observed with MERLIN in Chapman et al. (1994). They found, as
indicated in their figure 10, most of the red-shifted maser features
to occur in a large region $\approx 400 - 600$~mas from the most
blue-shifted features. However, they observed the bright red-shifted
features mainly in the 1665~MHz line and the bright blue-shifted
features at 1667~MHz. The maps of the 1665 and 1667~MHz transitions
were then aligned using the position of the centroids of the maser
emission at the most blue-shifted and red-shifted velocity. As a
result there could be a large error on the relative 1665 and 1667
maser positions. The position angle of the red-shifted features with
respect to the blue-shifted feature however, is the same as that of
our bright maser spot with respect to the transposed stellar
position. And since the separations are comparable it is likely that
the stellar position coincides with the most blue-shifted features
detected by Chapman et al. (1994). Although single dish spectra show
that the red-shifted side of the spectrum is $\approx 2.5$ times
brighter than the blue-shifted side, it is somewhat surprising that no
bright blue-shifted emission was detected in our observations. Either
the stellar image is too faint to observe at VLBA baselines or the
maser conditions changed between the epochs of observation (the MERLIN
observations were performed in 1986). This is quite possible since the
R~Cas masers are found to be very a-spherical, possibly due to the
presence of a companion star at $27.8$~arcsec separation (Proust et
al. 1981).

\subsubsection{S~CrB}

 The phase referencing of the S~CrB masers was performed on reference
source J1522+3144, at a separation of $0.4^\circ$. Because the source -- reference source
separation is so small, the errors due to the ionosphere are estimated to be less than
$2.5$ mas.

 The motion of S~CrB was also determined from observations of a
red-shifted maser feature. This feature seems to correspond to the
bright red-shifted maser emission observed in the single dish spectra
of both main-line maser transitions (e.g. Fix 1978, Etoka et
al. 2001). The brightest feature was detected in both transitions that
coincide within $\approx 5$~mas at each epoch, well within the beam
size. However, this feature cannot be the stellar image. Since there
are no published results on the morphology of the OH masers around
S~CrB we can only make a crude estimate on the systematic motion that
might be present for the red-shifted maser spot. With an outflow
velocity estimated from the single dish spectra of $\approx 4$~km~s$^{-1}$
and a stellar velocity of $0.0$~km~s$^{-1}$, a red-shifted feature at a
radial velocity of $~3$~km~s$^{-1}$ will have a transverse velocity of
$\approx 2.6$~km~s$^{-1}$. At the distance of S~CrB this would lead to a
systematic motion of $\approx 1.3$~mas/yr. Thus, a systematic error of
$\approx 0.9$~mas/yr could be present in the determined components of
the proper motion. Turbulence could cause random motions of $\approx
0.5$~mas/yr, while the phase referencing errors were estimated to be
$\approx 2.5$~mas. These errors however, were based on the separation
between both of the reference sources, while the separation between
S~CrB and the reference source used for the calibration is much
smaller. Thus, the residuals from our best fitted model ($\approx
1.2$~mas) are in excellent agreement with the phase referencing
errors, and the maser spot does not seem to show any unexplained
motions.

\subsection{The nature of the brightest maser features}

Whereas in U~Her and W~Hya the most blue-shifted maser spot is the
amplified stellar image, both R~Cas and S~CrB did not show any compact
blue-shifted maser spot.  Instead we observed a compact red-shifted
spot which could be traced for over 2 years. And even though it is not
the amplified stellar image, the bright red-shifted maser feature of
S~CrB coincides in the 1665 and 1667~MHz maser transition. Also, for
both these stars the single dish maser spectra show stronger
red-shifted emission than blue-shifted emission (e.g. Chapman et
al. 1994; Etoka \& Le Squeren 2000; Etoka et al. 2001).

As observed in one of the epochs of U~Her discussed in vL00, features
other than the most blue-shifted maser spot are occasionally dominant.
This is likely caused by high density maser blobs or filaments, which
can occur throughout the entire maser shell. Thus, these form of density
enhancements can also explain the occurrence of compact red-shifted
maser features.

In the case of R~Cas, the observations of Chapman et al.\
(1994), do show some compact emission from the most blue-shifted side
of the maser shell.  According to the comparison with our observations
this emission was likely the amplified stellar image.  However, it was
much weaker than the red-shifted emission and was not detected in our
observations.

\begin{table*}
\caption{Distances}
\begin{tabular}{|l||c|c|c|c|c|c|c|c|}
\hline
{ Source } & {  $K_0$ } & {  Hipparcos } & Hipparcos (1) & {  P--L (2) } & {  P--L (3) } & {  P--L (4) } & { 
P--L (5) } & {  VLBI } \\
  & {  (mag) } & {  (pc) } & (pc) & {
    (pc) } & {  (pc) } & {  (pc) } &
  {  (pc) } & {  (pc)} \\
\hline
\hline
{W~Hya } & {  -3.16 } & {  115 } & 78 &
  {  90 } & {  80 } & {  70 } &
  {  90 } & {  98$^{+30}_{-18}$} \\[3pt]
{S~CrB } & {   0.32 } & {  526 } & 417 &
  {  470 } & {  410 } & {  340
  } & {  450 } & {  433$^{+72}_{-54}$} \\[3pt]
{U~Her } & {  -0.29 } & {  610 } & 532 &
  {  380 } & {  340 } & {  265
  } & {  370 } & {  277$^{+112}_{-62}$} \\[3pt]
{R~Cas } & {  -1.80 } & {  106 } & 100 &
  {  200 } & {  170 } & {  135
  } & {  190 } & {  176$^{+92}_{-45}$} \\[3pt]
\hline
\multicolumn{9}{l}{(1) Knapp et al. (2003), (2) Whitelock \& Feast (2000), (3) Alvarez \&
  Mennessier (1997),}\\
\multicolumn{9}{l}{(4) Bedding \& Zijlstra (1998), (5) Chapman et al. (1994)}\\
\end{tabular}
\label{plsam}
\end{table*}

\begin{figure}[ht]
\begin{center}
   \resizebox{\hsize}{!}{\includegraphics{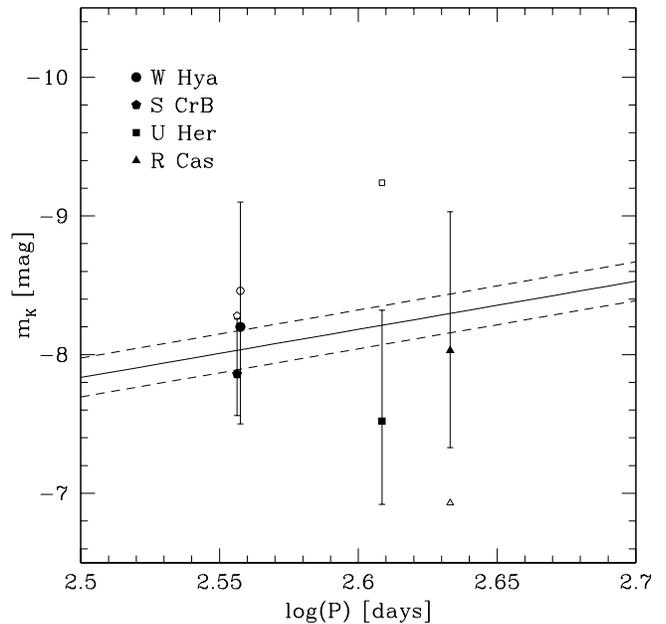}} \hfill
\end{center}
   \caption{Period vs. K$_0$ luminosity for the observed stars. The solid
   symbols and the error bars are determined using the distances
   obtained in this paper. The open symbols are the values using the
   Hipparcos distances. The solid line is the P-L relation determined
   by Whitelock \& Feast (2000) on the oxygen rich Mira stars observed
   with Hipparcos, the dashed lines are the spread in the relation due
   to the error in the P-L relation zero-point.}  \label{pl}
\end{figure}

\subsection{Distances}

 In Table~\ref{plsam} we compare the distances determined with VLBI
monitoring of the OH maser spots with those determined with a P-L
relation. The table also lists the Hipparcos distance (including
those as recalculated by Knapp et al. (2003), and the K$_0$ magnitude
of our sources. The P-L relation used to determine the values in the
column labeled (2) is also shown in Fig.~\ref{pl}. We use the P-L
relation of the form

\begin{equation}
M_K = -3.47 \log P + \beta,
\label{3_wf}
\end{equation}
with the slope of the relation derived from Large Magellanic Cloud
observation by Feast et al. (1989). The zero-point $\beta = 0.84 \pm
0.14$ was derived in Whitelock \& Feast (2000) from a sample of 180
oxygen rich Miras from the Hipparcos catalogue. We see in
Fig.~\ref{pl} that the luminosities obtained with the VLBI distances
show less scatter around the P--L relation than the Hipparcos points.

The distances were also compared with the P--L relation derived from
the Hipparcos data by Alvarez \& Mennessier (1997). They did not
a-priori assume a slope for the P--L relation and divided their
stellar sample into two groups based on the kinematics. For the first
group, which is thought to mainly consist of late disk stars they
found $M_K = - 3.41 \log P + 0.976$. For the second group, thought to
consist of stars that belong to an extended disk or to the halo, they
found $M_K = - 3.18 \log P - 0.129$. Based on their criteria it is
likely that all our sources are from the first group. The distance
values based on this relation are listed in the column labeled (3) of
Table~\ref{plsam}.

A third comparison was made with the P--L relation found by Bedding \&
Zijlstra (1998) for Mira and Semiregular (SR) stars from the Hipparcos
catalogue. They also did not fix the slope of the relation and found
$M_K = -1.67 \log P - 3.05$. They used a sample of 6 Miras and 18 SR
stars which had an Hipparcos parallax precision of $\sigma_\pi / \pi
\le 0.2$. The distance values based on this relation are listed in the
column labeled (4) of Table~\ref{plsam}.

It is of course possible, that since the Whitelock \& Feast, Alvarez
\& Mennessier and Bedding \& Zijlstra P--L relations are all based on
the Hipparcos data, a common bias could have been introduced. For
comparison we have also used the relation from Chapman et al. (1994),
which is given by Eq.\ref{3_wf} with $\beta=0.93$. This is solely
based on the LMC data of Feast et al. (1989), assuming a distance
modulus to the LMC of $18.55$. The resulting distances are shown in
the column labeled (5) in Table~\ref{plsam}.

The VLBI distances seem to agree best with the P--L relations from
Chapman et al. and Alvarez \& Mennessier with only U~Her being
systematically fainter.  As discussed in vL00, the size of U~Her
indicates that it pulsates in fundamental mode.  R~Cas is also
thought to be a fundamental mode pulsator as shown in van~Leeuwen
(1997). This was determined by observations of the stellar radius,
which seemed to be smaller than expected for overtone
pulsators. However, our revised distance implies a larger radius,
indicating that R~Cas is likely pulsating in an overtone. Because the
other stars have periods less than 400 days they are also expected to
be overtone pulsators. Since fundamental mode pulsators are somewhat
fainter than predicted by the P-L relation, this can explain the
off-set observed for U~Her.

\subsection{Perspectives on maser astrometry}

The previous result on U~Her (vL00) indicated that one could
follow the stellar trajectory accurately by VLBI monitoring the most
compact blue-shifted emission in OH masers. The amplification of the
stellar image would fulfill the requirement that an assumption has to
be made that relates the maser motion to the stellar motion. In
particular, to measure a parallax this tie must be much more accurate
than 1 AU. From the small sample presented here, it has emerged that
we cannot find such a special spot in every stellar OH maser. However,
even without a stellar image, valid results can still be obtained from
OH astrometry. For R~Cas and S~CrB we have successfully traced a
bright maser spot in the shell that cannot be the amplified stellar
image, as it originates from the far side of the expanding shell. The
parallax can still be measured if we can assume that this maser spot
has a constant relative motion with respect to the star. The proper
motion of the star can only be obtained if the maser's peculiar motion
can be assumed to be small or accurately known, for instance from the
geometry of the shell. Our results indicate that the maser's motion
with respect to the star is indeed small. Also, it is clear that these
maser spots are persistent on the timescale of a few years.

The accuracy of the current measurements is limited by the ionosphere
and the extrapolation of the calibrator phases. In principle, this
restriction can be addressed by ionospheric calibration or by using
closer (in-beam) position calibrators, as was shown for VLBI pulsar
astrometry by Fomalont et al. (1999). However, for the OH masers, a 1
mas accuracy limitation seems to be intrinsic, as the maser spot
brightness is relatively modest. A high precision parallax can be
obtained by using many observation epochs and the proper motion can
benefit from extending the time baseline, provided individual maser
spots last long enough. At the OH maser frequency the technique is
relatively straightforward, as the coherence time is long and
calibrators are abundant.  However, in this way, parallaxes can only
be obtained for sources closer than 1~Kpc. For SiO and H$_2$O masers
the restrictions set by the limited brightness are lifted and stars
much further away could be probed. However, the coherence times are
shorter and calibrators will be harder to find.

\section{Conclusions}

We have improved upon the distances and proper motions of U~Her, S~CrB
and R~Cas using VLBI astrometry of the 1665 and 1667~MHz OH masers.
Also for the relatively close star W~Hya we have fitted a stellar
trajectory. However, large residuals remain in this fit and we
attribute these to variations in the stellar photosphere which are
important for a star as close as W~Hya (\S~\ref{whyasec}). These
variations introduce significant additional motions to a maser
spot which is the amplified stellar image.

For U~Her we traced the amplified stellar image, which we have now
been able to follow for almost 8 years. For W~Hya it is plausible that
the bright maser spot is also the amplified stellar image. The other
two stars do not show any blue maser spots that can be detected with
VLBI, implying that the amplified stellar image is not dominant in all
OH maser shells. Furthermore, it has been observed that the 1665 MHz
and 1667 MHz masers can coincide on VLBI scales, even when they are
not the amplified stellar image. Still, even the red-shifted
maser spots are shown to allow the determination of a good VLBI
parallax and proper motion.

For phase referencing at 1.6 GHz the limitation in accuracy is the
ionosphere, which is clearly demonstrated by the fact that the data
quality steeply degrades near the maximum of solar activity.  In all
cases, but W~Hya, the residuals in the data with respect to the best
fit, can be entirely attributed to the ionosphere. Thus, there are no
indications of additional motions in these stars, like, for instance,
could be expected in binary systems.

We have shown that astrometry of OH maser spots with VLBI offers a
unique possibility to measure the distances of enshrouded AGB
stars. The four stars stars studied here are presumably losing mass
more rapidly than the bulk of Mira variables, mostly studied with
Hipparcos. Even so, we find these four stars now closely follow the
established $P-L$ relation and fit the recent theories on pulsation
modes.

{\it acknowledgments:} We thank Sandra Etoka for kindly providing the
  single dish maser spectra of U~Her, W~Hya and S~CrB. Furthermore, we
  acknowledge the use of the JIVE computing facilities for handling
  the large (50 GB) data set and thank the JIVE support staff for
  their help. WV thanks the Niels Stensen Foundation for partly
  supporting his stay at Cornell University.

\end{document}